\documentstyle[12pt]{article}
\def\theequation{\arabic{equation}}

\renewcommand{\thefootnote}{\fnsymbol{footnote}}
\newcommand{\r}[1]{(\ref{#1})}
\begin{document}
\thispagestyle{empty}

\begin{center}
{\large{\bf TRACE ANOMALIES AND  COCYCLES
OF WEYL AND DIFFEOMORPHISMS GROUPS}} \vspace{1.5cm}\\

{\large D.R. Karakhanyan\footnote{E-MAIL:karakhan@tpd.
erphy.armenia.su},}\vspace{0.5cm}
{\large R.P.Manvelyan\footnote{E-MAIL:manvel@tpd.
erphy.armenia.su }}\vspace{0.5cm}\\
and\vspace{0.5cm}\\
{\large  R.L. Mkrtchyan \footnote{ E-MAIL mrl@dircom.erphy.
armenia.su}}\vspace{0.5cm}\\

{\it Theoretical Physics Department,}\\
{\it Yerevan Physics Institute}\\
{\it Alikhanyan Br. st.2, Yerevan, 375036 Armenia }
\end{center}
\bigskip
\bigskip
\bigskip
\bigskip
\begin{abstract}
The general structure of trace anomaly, suggested recently by
Deser and Shwimmer, is argued to be the consequence of the Wess-Zumino
consistency condition. The response of partition function on a finite
Weyl transformation, which is connected with the cocycles of the Weyl
group in $d=2k$ dimensions is considered, and explicit answers for
$d=4,6$ are obtained. Particularly, it is shown, that addition of the
special combination of the local counterterms leads to the simple form
of that cocycle, quadratic over Weyl field $\sigma$, i.e. the form,
similar to the two-dimensional Lioville action. This form also
establishes the connection of the cocycles with conformal-invariant
operators of order $d$ and zero weight. Beside that, the general rule
for transformation of that cocycles into the cocycles of
diffeomorphisms group is presented.
\end{abstract}

\vfill
\setcounter{page}0
\renewcommand{\thefootnote}{\arabic{footnote}}
\setcounter{footnote}0
\renewcommand{\theequation}{\arabic{section}.\arabic{equation}}
\setcounter{equation}{0}
\newpage

\newcommand{\ra}{\rightarrow}

\pagestyle{plain}

\section{\bf Introduction.}
\setcounter{equation}{0}

        The conformal, or Weyl, or trace anomaly, discovered more than
20 years ago
\cite{Duff} plays an important role in understanding of many phenomena.
It helps in the calculation of the effective action of 2D gravity
\cite{Pol}, in understanding of the properties of gauge and supersymmetric
gauge theories,  and in other places. Despite the much
efforts, aimed on the understanding of the structure of gauge anomalies
at the beginning of 80-s, the structure of the trace anomaly remains
unclerified. The understanding of that structure appeared at 1993 due to the
work of
Deser and Schwimmer \cite{DS}, where they investigated a
structure of Weyl anomaly using a scailing property of effective action.
Beside that
in the ref.\cite{BN}  the new approach  of obtaining of
sigma-model-like
generating functional (like in ref.\cite{Wit} for gravitational and gauge
anomaly) for conformal anomaly in high dimension was elaborated.
But both of these approaches
don't lead to manifest expressions in the dimension $d>4$.
Our approach of investigation of conformal anomaly in higher even dimensions is
based on the well-known criterion for definition of possible forms of
anomalies - on the
Wess-Zumino consistensy condition \cite{WZ} , which in this abelian case
concides
with the property of symmetricity of the conformal variation of anomaly.
In this article we consider the general structure of trace anomaly
in all dimensions and argue, that the structure, suggested by Deser and
Schwimmer \cite{DS}, is actually the consequence of Wess-Zumino consistency
condition, although we were not able to present the complete proof.
Then we investigate the finite form of the trace anomaly - i.e the change
of the measure in the functional integral under finite Weyl transformations.
It is well-known \cite{FS}, that corresponding factor has the properties of
the 1-cocycle - of the Weyl group, in this case.
The analysis of the situation in  d=4 (\cite{Rieg}) and d=6 (present work, see
next section) leads to the beautiful relation between structure of anomaly,
1-cocycles of Weyl group, Euler classes and zero weight
conformal-invariant operators of $d$-th order
in $d=2k$ dimensions. We found, that the non-local effective action,
wich  generates the trace anomaly, has a form of Polyakov's
non-local action in d=2
with  $d$-th order conformal invariant operator instead of Laplacian and
(essentially) the density of Euler characteristik instead of scalar curvature.
It means that corresponding nontrivial cocycle of Weyl group
is a second-order local, Weyl-invariant functional on group
parameter $\sigma (x)$ (see Sect.3).
It is shown below, that generally, all the cocycles are connected
with Weyl-invariant local Lagrangians - polynomials over derivatives of
$\sigma(x)$, but all such Lagrangians of order higher than two
give rise to the trivial cocycles, i.e. coboundaries.
Beside that there are linear nontrivial
cocycles connected with conformal-invariant scalar combination of curvature
tensor and its derivativs.

We investigate also in general form the well-known
possibility of transforming the Weyl anomaly through local counterterms in
effective
action into the anomaly of diffeomorphisms group, and obtain the general
prescription for that mapping (the resulting diffeomorphisms anomaly actually
violates diffeomorphisms group only partially, maintaining the
volume-preserving part of that group\cite{RRD}, ($2d$ case see in \cite{KMM})).
 Difference with $2d$ case lies in a fact, that
   at $d>2$ the Weyl anomaly is not enough for construction of full effective
action $W(g_{\mu\nu})$ of conformal matter field, because the parameters of
local symmetry group:  $$ Weyl \otimes  Diff(2k) $$ do not cover all
components of the metric.  But one has a possibility of construction of the
finite variation of effective action on Weyl rescaling of metric:  $$ S(\sigma
,g_{\mu\nu}) = W(e^{\sigma} g_{\mu\nu}) - W(g_{\mu\nu})$$. This local action
corresponds in $d=2$ to Liouville action \cite{Pol}. This action $ S(\sigma,
g_{\mu\nu})$ has a property of being 1-cocycle of Weyl group, and may be used
for transition from $ W(g_{\mu\nu})$ to Weyl invariant effective action
$ \tilde W(g_{\mu\nu})$, the finite variation of which under the diffeomorphism
$x^{\mu} \ra f^{\mu}(x)$ gives us the 1-cocycle of diffeomorphisms group
$\tilde S(f^{\alpha},g_{\mu\nu})$, ($2d$ case see in \cite{KMM}).
All these statments we have proved for $d=4,6$ by explicit calculations,
and they are the basis of our hypothesis in higher dimensions, together with
some additional arguments.
 The organization of the paper is as follows. In section 2 we consider
 the connection between Wess-Zumino consistency condition
\cite{WZ} and structure of trace anomaly .In section 3 we construct all
$1$-cocycles of Weyl group in $d=4,6$ and show the connection between
nontrivial cocycles and conformal-invariant operators.
In section 4 we define the local counterterms for
transition from Weyl to $Diff(d)$ invariant effective action.
In conclusion the results and perspectives are summarised.

\newpage

\section{\bf Weyl anomaly and Wess-Zumino consistency condition}
\setcounter{equation}{0}

Let's consider the effective action  for conformal matter $\varphi$
 in  external gravitational field:
\begin{equation}
\label{1}
W(g) = \ln \int D_{g}\varphi \exp\{ -S_{w}(\varphi; g)\}
\end{equation}
where $ S_{w}(\varphi;g)$ is classical Weyl and diffeomorphism invariant
action for matter fields and where Weyl transformation is defined as:
\begin{equation}
\label{2}
g_{\alpha\beta} \ra e^{\sigma(x)} g_{\alpha\beta};
 \qquad  \varphi \ra e^{\Delta\sigma(x)}\varphi
\end{equation}
Where $\Delta$ is conformal weight of matter field.

Then, taking an infinitesimal $\sigma$, we can write down the equation
for anomaly
\begin{equation}\label{3}
\delta_{\sigma} W(g) = \int T^{\mu}_{\mu} \sigma(x) \sqrt{g} d^{2k}x
\end{equation}
The Wess-Zumino consistency condition in the the case of Weyl transformations
is simply a statement of a symmetricity of second conformal
 variation of effective action:
\begin{equation}\label{4}
\frac{\delta ^{2}W(g)}{\delta\sigma(x)\sqrt{g}\delta\sigma(y)} =
\frac{\delta ^{2}W(g)}{\delta\sigma(y)\sqrt{g}\delta\sigma(x)}
\end{equation}
 or, in other words
\begin{equation}\label{5}
\frac{\delta A(x)}{\delta\sigma(y)} =\frac{\delta A(y)}{\delta\sigma(x)}
\end{equation}
where we denote  $T^{\mu}_{\mu}=A(x)$

Then we can propose the following hypothesis about the structure of solution
of Wess-Zumino consistency condition in all (even) dimensions.
Let we have local function
of metric $A(g)$ i.e. the anomaly. The requirement, that the conformal
variation of that function is symmetric (WZ consistency condition) leads
to the following statement, concerning the structure of $A(g)$:
1) $A(g)$ is the sum with an arbitrary
coefficients of the following terms.\\
a) Euler density.\\
b) Weyl-invariant polynomials over Riemann tensor and it's covariant
derivatives.\\
c) Covariant total derivatives of polynomials over Riemann tensor and
it's covariant derivatives.\\
In addition, we are making also the following statement
2) Terms c) are the Weyl variations of local functionals of metric.
Taking into account the fact, that the definition of the measure in the
functional integral always can be changed by multiplying on the exponent
of the local functionals (counterterms) of metric, one deduce,
that third type (i.e. c) type)
solutions of WZ condition are in that sense inessential and will be
called trivial below.

This hypothesis agree with results obtained earlier in $d=2,4$ \cite{Duff}
and agree with our investigation in $d=6$. This allow us to be suggest
these statements also in $d>6$.

Let's consider the solution of equation \r{5}, and, correspondingly, test
the validity of the above hypothesis, in some special cases, when anomaly
$A(x)$ satisfies some additional constraints.

1.In the case when $A$ is local polynomial function of Riemann tensor
$R^{\alpha\beta}_{\mu\nu}$ , eq.\r{5} gives the condition
\begin{equation}\label{6}
\nabla^{\alpha}\frac{\delta A(R)}{\delta R^{\alpha}_{\mu\beta\nu}}
g_{\mu\nu} = 0
\end{equation}
Then we have two types of solutions of \r{6}:
First, Weyl invariant $k$-th degree  polynomial of Weyl tensor
$C^{\alpha\beta}_{\mu\nu}$, and, next,  (Weyl noninvariant) Euler density:
\begin{equation}\label{7}
E_{2k} =\frac{1}{2k!}
\varepsilon^{\mu_{1}\nu_{1}\mu_{2}\nu_{2}\cdots\mu_{k}\nu_{k}}
\varepsilon_{\alpha_{1}\beta_{1}\alpha_{2}\beta_{2}\cdots\alpha_{k}\beta_{k}}
R^{\alpha_{1}\beta_{1}}_{\mu_{1}\nu_{1}}
R^{\alpha_{2}\beta_{2}}_{\mu_{2}\nu_{2}}\cdots
R^{\alpha_{k}\beta_{k}}_{\mu_{k}\nu_{k}}
\end{equation}
These are the solutions of type a) and b).

2.In the case when $A$ is total derivative of the form:
\begin{equation}\label{8}
A(x)=\nabla_\alpha\nabla_\beta V^{\alpha\beta}(R)
\end{equation}
where $V^{\alpha\beta}(R)$ is local polynomial $k-1$ degree on $R$, we obtain
the following restriction :
\begin{equation}\label{9}
\frac{\delta V^{\alpha\beta}(R)}{\delta R^{\lambda}_{\mu\rho\nu}}
\delta^{\lambda}_{\rho} =  \frac{\delta V^{\mu\nu}(R)}
{\delta R^{\lambda}_{\alpha\rho\beta}}\delta^{\lambda}_{\rho}
\end{equation}
 The solutions of type \r{8}, which satisfy \r{9}, coincide with
 variations of all independent local counterterms to effective action.
 For example in $d=4$ one has only one appropriate counterterm, which is
 a second order in curvature,
 with independent conformal variation:
\begin{equation}\label{10}
W_0 = \int R^2 \sqrt{g} d^{4}x ;
\qquad \frac{\delta S_{0}}{\delta\sigma} = -6\sqrt{g} \Box R
\end{equation}
Two other possible counterterms
\begin{equation}\label{11}
   \int R_{\mu\alpha\nu\beta}R^{\mu\alpha\nu\beta}\sqrt{g}d^{4}x;
   \qquad\int R_{\mu\nu}R^{\mu\nu}\sqrt{g}d^{4}x;
\end{equation}
 give the same \r{10} contribution to anomaly. That can be checked
 directly, but also follows from the fact, that there are two
 independent constraints
on the variation of these three counterterms. These constraints are
consequences of the Weyl-invariance of the actions with following densities:

1. $C_{\mu\alpha\nu\beta}C^{\mu\alpha\nu\beta}$

2.density of the Euler characteristic: $E_4$
Weyl variations of these actions give abovementioned constraints.

 From the other hand, the terms of the type \r{8} are in this actually
unique, and coincide with \r{10}, so the statements 1) and 2) are proved
in this (i.e. $d=4$) case.

The similar situation is in $d=6$.
Let's consider all third-order combinations of $ C_{\mu\alpha\nu\beta},
R_{\mu\nu}, R$ and Laplace operator - i.e. all six-dimensional dimensionless
actions.
\begin{equation}\label{12}
A_1 = C^{\alpha\beta\mu\nu}C_{\alpha\beta\lambda\rho}C^{\lambda\rho}_{\mu\nu}
\end{equation}
\begin{equation}\label{13}
A_2=C^{\alpha\mu\beta\nu}C_{\alpha\rho\beta\lambda}
{{{C^{\rho}}_{\mu}}^{\lambda}}_{\nu}
\end{equation}
\begin{eqnarray}\label{14}
L_1 & = & R^{3}\nonumber\\
L_2 & = & R^{\alpha\beta}R_{\alpha\beta}R\nonumber\\
L_3 & = & C^{\alpha\mu\beta\nu}R_{\alpha\beta}R_{\mu\nu}\\
L_4 & = & C^{\alpha\mu\nu\rho}C_{\beta\mu\nu\rho}R^{\beta}_{\alpha}\nonumber\\
L_5 & = & C^{\alpha\beta\mu\nu}C_{\alpha\beta\mu\nu}R\nonumber\\
L_6 & = & R \Box R \nonumber
\end{eqnarray}
\begin{eqnarray}
L'_2 & = & R^{\alpha\mu}R_{\mu\beta}R^{\beta}_{\alpha}\label{eq:15}\\
L'_6 & = & R_{\alpha\beta} \Box R^{\alpha\beta}\label{eq:16}\\
L''_6 & = & C^{\alpha\beta\mu\nu} \Box C_{\alpha\beta\mu\nu}\label{eq:17}
\end{eqnarray}

 All other non-fullderivative combinations coincide with the combination
of this Lagrangians due to the Bianchi identities.
It's easy to see that first two densities are Weyl-invariant and can be
considered as a contribution in anomaly from nonlocal part of effective
action. Beside that, one of these densities (say, $L'_2$), has dependent
conformal variation due to the Gauss-Bonnet theorem.
The counterterms  constructed from last two densities have dependent conformal
variation, since there are two  additional conformal-invariant
actions with following densities:
\begin{eqnarray}\label{18}
A_{3} & = & C^{\alpha\beta\mu\nu} \Box C_{\alpha\beta\mu\nu}
+2C^{\alpha\mu\nu\rho}C_{\beta\mu\nu\rho}R^{\beta}_{\alpha}\nonumber\\
      &   & -3C^{\alpha\mu\beta\nu}R_{\alpha\beta}R_{\mu\nu}
      -\frac{3}{2}R^{\alpha\mu}R_{\mu\beta}R^{\beta}_{\alpha}\\
  &   & +\frac{27}{20}R^{\alpha\beta}R_{\alpha\beta}R -
  \frac{21}{100}R^3\nonumber
\end{eqnarray}
\begin{eqnarray}\label{19}
 A_4 & = &R_{\alpha\beta} \Box R^{\alpha\beta}-\frac{3}{10}R \Box R\nonumber\\
&   & -2C^{\alpha\mu\beta\nu}R_{\alpha\beta}R_{\mu\nu}
 +R^{\alpha\mu}R_{\mu\beta}R^{\beta}_{\alpha}\\
&   & - \frac{1}{10}R^{\alpha\beta}R_{\alpha\beta}R - \frac{1}{50}R^3\nonumber
\end{eqnarray}
They differ from $A_{1,2}$ in that they have nonzero Weyl
variations
($\delta_{\sigma}A_{3,4} =$ full derivatives), therefore we can use they
only as constraint on local counterterms, but not as an independent
contributions into the anomaly.

Therefore in $d=6$ there are only 6 independent local counterterms
\begin{equation}\label{20}
W^{i}_o = \int L_i\sqrt{g}d^6x,\quad i=1,2,\cdots 6
\end{equation}
and 3 nontrivial contributions in anomaly $E_6,A_1,A_2$

Again, it is possible to show that all possible solutions of the WZ condition
\r{9} of the type \r{8} coincide with the combinations of the variations of
the actions $W^{i}$.

These considerations give the complete proof of statements 1) and 2) at d=6.

\section{\bf Cocycles and conformal-invariant operators}
\setcounter{equation}{0}

Let's consider the problem of the change of the measure in the functional
integral for conformal matter field $\varphi$ in external gravitational field
under the finite Weyl transformation \r{2}.

 The measure in the functional integral changes in the following way:
\begin{equation}\label{21}
  D_{e^{\sigma(x)}g}\varphi = D_{g} \varphi \exp{S(\sigma;g_{\alpha\beta})}
\end{equation}
This type of relations are very important, being the starting point of
DDK calculation of the critical exponent of $2d$ gravity \cite{DDK}.

 The action $S(\sigma; g)$ in \r{2} has to satisfy some conditions.
First,in the case of infinitisimal transformation $\delta\sigma(x)$
it has to reproduce the trace anomaly:
\begin{equation}\label{22}
S(\delta\sigma(x); g_{\alpha\beta}) =
\int T^{\mu}_{\mu} \delta\sigma(x) \sqrt{g} d^{2k}x
\end{equation}
Second,  $S(\sigma; g)$  has to satisfy the following property, which
follows from the application of \r{2} to the composition of two Weyl
transformations $\sigma_{1}$ and $\sigma_{2}$:
\begin{equation}\label{23}
S(\sigma_{1} + \sigma_{2}; g) = S(\sigma_{1}; e^{\sigma_{2}}g)
+  S(\sigma_{2}; g)
\end{equation}
which means that  $S(\sigma; g)$  is the 1-cocycle of the group of Weyl
transformations \cite{FS}.

On the other hand, the action  $S(\sigma; g)$ coincides with the finite
variation of anomalous effective action, due to the propertie \r{1} and
\r{21}.
In other words
\begin{equation}\label{24}
S(\sigma, g) =  W(e^{\sigma} g) - W(g)
\end{equation}
and non-triviality of the cocycle $S(\sigma;g)$  follows from the fact that
$W(g)$ is non-local, $Diff(2k)$-invariant  functional of $g_{\alpha\beta}$,
 Thus, we can easily calculate the trivial cocycles as a coboundary of
 local counterterms   $W^{i}_{0}(g)$ which we shall call from now on
 0-cochains:
\begin{equation}\label{25}
S^{i}_{0}(\sigma,g) = \triangle W^{i}_{0}(g)
\end{equation}
where we have defined coboundary operator $\triangle$ on 0-cochains
as finite Weyl variation \r{24}.
Then we can define 1-cochains as a local functions $W_{1}(\sigma,g)$ of
group parameter and metric with coboundary operator:
\begin{equation}\label{26}
\triangle W_{1}(\sigma_{1},\sigma_{2},g) = W_{1}(\sigma_{1} + \sigma_{2}; g)
 - W_{1}(\sigma_{1}; e^{\sigma_{2}}g) -  W_{1}(\sigma_{2}; g)
\end{equation}
It is easy to see that $\triangle^2 = 0$ which is exactly
the cocyclic
property \r{23}. One can generalize this
construction on higher cohomologies of Weyl group \cite{FS}.

The nontrivial cocycles can be obtained from  the solution of eq. \r{23}.
If we are looking for the general solution of equation \r{23} with
condition \r{22}
we have to take $\sigma_2 =\sigma$ and $\sigma_1 =\delta\sigma$ and get
the differential form of \r{29}:
\begin{equation}\label{27}
 \delta S(\sigma; g) = S(\delta\sigma; e^{\sigma}g)
= \int A(R(e^{\sigma}g))\delta\sigma \sqrt{g} d^{2k}x
\end{equation}
The explicit form of solution for two-dimensional case is
the famous Liouvill action \cite{Pol}
\begin{equation}\label{28}
S_{d=2}(\sigma,g)=\frac{c}{48\pi}\int d^2 x\sqrt{g}(\frac{1}{2}g^{\alpha
\beta}\partial_\alpha \sigma \partial_\beta \sigma + R \sigma)
\end{equation}

In four dimensions explicit form of cocycle, corresponding to $E_4$, firstly
has been found in \cite{Rieg}
\begin{eqnarray}\label{29}
S_{E}(\sigma,g)&=&\int d^4 x\sqrt{g}\frac{1}{(2\pi)^2}(\{\frac{1}{8}
[(\nabla_\alpha \sigma \nabla^\alpha \sigma)^2
+\frac{1}{2}\nabla_\alpha \sigma
\nabla^\alpha \sigma \nabla^2 \sigma\nonumber\\ &
&-(R^{\alpha\beta}-\frac{1}{2}g^{\alpha\beta}R)\nabla_\alpha \sigma
\nabla_\beta \sigma]\} +\sigma E_4)
\end{eqnarray}
In $d=4$ there is linear nontrivial cocycle corresponding to single invariant
density $C_{\mu\alpha\nu\beta}C^{\mu\alpha\nu\beta}$:
\begin{equation}\label{30}
S_{C}(\sigma,g) = \int C_{\mu\alpha\nu\beta}C^{\mu\alpha\nu\beta}\sigma (x)
\sqrt{g}d^4x
\end{equation}
This expression satisfies the cocyclic property \r{23} and can appear
in the Weyl transformation of the measure \r{21}.

Then in d=4 we can obtain only one trivial cocycle:
\begin{equation}\label{31}
S_{0}(\sigma,g) = \triangle\int R^2\sqrt{g}d^4x
\end{equation}
On the other side, one can check that
\begin{equation}\label{32}
 {\bf S}(\sigma,g) = S(\sigma,g) + \frac{2}{3} S_{0}(\sigma,g) =
\int[\frac{1}{2}\sigma \Delta_4 \sigma + (E_4 +
\frac{2}{3}\Box R)\sigma]\sqrt{g}d^4x
\end{equation}
where
\begin{equation}\label{33}
\sqrt{g}\Delta_4 =\sqrt{g}( \Box^2 - 2R^{\mu\nu}\nabla_{\mu}\nabla_{\nu}
+ \frac{2}{3}R\Box - \frac{1}{3}(\nabla^{\mu} R)\nabla_{\mu})
\end{equation}
is the fourth-order conformal-invariant differential operator acting on a
scalar
field of zero conformal weight \cite{Rieg} (exact analogy of $\sqrt{g}\Box$
in$d=2$).

Thus nontrivial cocycle in $d=4$ can be cut up to the second order on $\sigma$
by adding a  trivial cocycle with appropriate coefficient.
The reason of this reduction comes from the following observation, which
also explains the Weyl-invariance property of the operator \r{33}.
It is easy to see that from cocyclic property \r{23} follows that the
highest-order over $\nabla\sigma$ term in cocycle has to be Weyl-invariant.
  There is only
one fourth-order invariant action in $d=4$
\begin{equation}\label{34}
I_4 = \int d^4x\sqrt{g}(\nabla_\mu\sigma\nabla^\mu\sigma)^2
\end{equation}
and there is no third-order invariant action in $d=4$,
consequently we can use trivial
cocycle \r{31} for the reduction of nontrivial one  to the
second-order cocycle
\r{32}. The corresponding highest order part involves Weyl-invariant
operator \r{33}.
These considerations leads to the correspondence between cocycles and
Weyl-invariant actions,
moreover, nontrivial cocycle is connected to the second-order
invariant action and
conformal invariant fourth-order differential operator.

Let's now consider the case $d=6$.
Firstly we can easily construct two linear cocycles using invariant part
of anomaly:
\begin{equation}\label{35}
S^{1}_C =\int\sqrt{g}d^6x C^{\alpha\beta\mu\nu}C_{\alpha\beta\lambda\rho}
C^{\lambda\rho}_{\mu\nu}\sigma (x)
\end{equation}
\begin{equation}\label{36}
S^{1}_C =\int\sqrt{g}d^6x C^{\alpha\mu\beta\nu}C_{\alpha\rho\beta\lambda}
{{{C^{\rho}}_{\mu}}^{\lambda}}_{\nu}\sigma (x)
\end{equation}
Secondly let's obtain the solution of equation (27) with $A(g) = E_6$:
 \begin{eqnarray}\label{37}
S_E & = & \int\sqrt{g}d^6x\{-6(\nabla^{\mu}\sigma\nabla_{\mu}\sigma)^{3}
-18 \Box \sigma (\nabla^{\mu}\sigma\nabla_{\mu}\sigma)^{2}\nonumber\\
&   & +24 \nabla^{\lambda}\sigma\nabla_{\lambda}\sigma[\nabla^{\mu}
\nabla^{\nu}\sigma\nabla_{\mu}\nabla_{\nu}\sigma -(\Box \sigma)^2]
 - 6R(\nabla^{\mu}\sigma\nabla_{\mu}\sigma)^{2}\\
&   & -4G^{\lambda\rho}_{\mu\nu}\nabla_{\lambda}\sigma\nabla^{\mu}\sigma
\nabla^{\nu}\nabla_{\rho}\sigma
 -3G^{\nu}_{\mu}\nabla^{\mu}\sigma\nabla_{\nu}\sigma +
6!E_{6}\sigma\}\nonumber
\end{eqnarray}
where
\begin{eqnarray}\label{38}
G^{\lambda\rho}_{\mu\nu} = \varepsilon^{\mu_1\nu_1\mu_2\nu_2\lambda\rho}
\varepsilon_{\alpha_1\beta_1\mu_2\nu_2\mu\nu}
R^{\alpha_1\beta_1}_{\mu_1\nu_1},\nonumber\\
G^{\nu}_{\mu} = \varepsilon^{\mu_1\nu_1\mu_2\nu_2\lambda\nu}
\varepsilon_{\alpha_1\beta_1\alpha_2\beta_2\lambda\mu}
R^{\alpha_1\beta_1}_{\mu_1\nu_1}R^{\alpha_2\beta_2}_{\mu_2\nu_2}
\end{eqnarray}
Then we can calculate 6 trivial cocycles $S^i_0$ using \r{14}, \r{20}
 and
\r{25}.
These cocycles correspond to all Weyl-invariant actions of order 3-6 over
$\nabla_{\mu}\sigma$. In other words, we can easily  check that:
\begin{eqnarray}\label{39}
 S_6 & = & \frac{1}{125}S^1_O = I_6 + ...\nonumber\\
S_4 & = & \frac{1}{125}S^1_0 +\frac{1}{25}S^6_0 = I_4 + ...\nonumber\\
S'_4 & = & \frac{1}{25}S^2_0 - \frac{1}{125}S^6_0 =I'_4 + ...\nonumber\\
S_3 & = & S^3_0 = I_3 + ...\\
S_2 & = & S^4_0 = I_2 + ...\nonumber\\
S'_2 & = & S^5_0 = I'_2 + ...\nonumber
\end{eqnarray}
where
\begin{eqnarray}\label{40}
I_6 & = & \int\sqrt{g}(\nabla^{\mu}\sigma\nabla_{\mu}\sigma)^3\nonumber\\
I_4 & = &\int\sqrt{g}(\nabla^{\mu}\sigma\nabla_{\mu}\sigma)[\Box +
\frac{1}{5}R]
(\nabla^{\mu}\sigma\nabla_{\mu}\sigma)\nonumber\\
I'_4 & = &\int\sqrt{g}[\nabla^{\mu}\nabla^{\nu}\sigma\nabla_{\mu}\nabla_{\nu}
\sigma\nabla^{\alpha}\sigma\nabla_{\alpha}\sigma\\
&   & - \frac{1}{2}\nabla^{\mu}
(\nabla^{\alpha}\sigma\nabla_{\alpha}\sigma)\nabla_{\mu}
(\nabla^{\beta}\sigma\nabla_{\beta}\sigma) - \frac{1}{4}(\Box\sigma)^2
(\nabla^{\alpha}\sigma\nabla_{\alpha}\sigma)]\nonumber\\
I_3 & = & \int\sqrt{g}C^{\alpha\mu\beta\nu}\nabla_{\alpha}\nabla_{\beta}\sigma
\nabla_{\mu}\nabla_{\nu}\sigma\nonumber\\
I_2 & = & \int\sqrt{g} C^{\alpha\mu\nu\rho}C_{\beta\mu\nu\rho}\nabla^{\beta}
\sigma\nabla_{\alpha}\sigma\nonumber\\
I'_2 & = & \int\sqrt{g} C^{\alpha\beta\mu\nu}C_{\alpha\beta\mu\nu}
\nabla_{\rho}\sigma\nabla^{\rho}\sigma\nonumber
\end{eqnarray}
 are conformal-invariant actions.
 Then we can obtain the main result of this section.The following nontrivial
 cocycle :
\begin{equation}\label{41}
 {\bf S}_2 = S_E +6S_6 +12S_4 - 60S'_4 -8S_3
\end{equation}
is second order on $\sigma$ and has a form
\begin{equation}\label{42}
{\bf S}_2 \sim \int\sqrt{g}\sigma\Delta_6\sigma + .....
\end{equation}
where $\Delta_6$ is Weyl-invariant  sixth-order differential operator
in $d=6$. Differently from $d=4$ case this operator is not unique, since
we can add
to the action ${\bf S}$ the second-order cocycles $S_2$ and $S'_2$
with arbitrary
coefficients, which leads to the arbitrariness in definition of $\Delta_6$.
\begin{eqnarray}\label{43}
\Delta_6 & = &\Box^3 -\frac{16}{3}\nabla_\alpha\nabla_\beta
C^{\alpha\mu\beta\nu}\nabla_\mu\nabla_\nu
-2\nabla_\alpha\nabla_\beta R^{\alpha\mu}\nabla^\beta\nabla_\mu\nonumber\\
&  &-2\nabla_\alpha\nabla_\beta R^{\beta\mu}\nabla^\alpha\nabla_\mu
 +\frac{2}{5}\nabla_\alpha\nabla_\beta R\nabla^\alpha\nabla^\beta
  +\frac{3}{5}\Box R\Box
\nonumber \\
&  &-\frac{8}{3}\nabla_\alpha C^{\alpha\mu\beta\nu}R_{\mu\nu}\nabla_\beta
-\nabla_\alpha R^{\alpha\mu}R_{\mu\beta}\nabla^\beta
+\frac{1}{2}\nabla^\alpha R^{\mu\nu}R_{\mu\nu}\nabla_\alpha\\
&  &+\frac{1}{5}\nabla_\alpha RR^{\alpha\beta}\nabla_\beta
-\frac{6}{25}\nabla^\alpha R^2 \nabla_\alpha
- \frac{2}{5}\nabla^{\mu}(\Box R)\nabla_\mu\nonumber\\
&  &+\varepsilon^{\mu_1\nu_1\mu_2\nu_2\lambda\nu}\varepsilon_
{\alpha_1\beta_1\alpha_2\beta_2\lambda\mu}R^{\alpha_1\beta_1}_{\mu_1\nu_1}
xR^{\alpha_2\beta_2}_{\mu_2\nu_2}\nabla^\mu\nabla_\nu\nonumber\\&  &
+ \theta\nabla_\alpha C^{\alpha\mu\nu\rho}C_{\beta\mu\nu\rho}\nabla^\beta
+\tau\nabla_\alpha C^{\beta\mu\nu\rho}C_{\beta\mu\nu\rho}\nabla^\alpha
\nonumber
\end{eqnarray}
where $\tau$ and $\theta$ arbitrary parameters.

The technical difficulties do not permit us to curry on the similar explicit
calculations in higher
dimensions, but these results convinced us, that in the same way
one can cut the non-trivial cocycle, corresponding to the Euler
characteristic term in the anomaly to the second order over $\nabla\sigma$
in all higher dimensions.

\section{\bf Cocycles of diffeomorphism group from the cocycles of the Weyl
group}
\setcounter{equation}{0}

Let's now consider the new non-local effective action:
\begin{equation}\label{44}
\tilde W(g) = W(g) + S(\sigma; g)\bigg\vert_{\sigma= - \frac{1}{k}\ln\sqrt{g}}
\end{equation}
where dimension of space-time is $d=2k$.
It's ease to see that due to the relation \r{23} (cocyclic property)
this action is Weyl-invariant, but not diffeomorphism-invariant.
Let's now in analogy with \r{24} calculate the finite
variation of $\tilde W(g)$
under diffeomorphism transformation:
\begin{equation}\label{45}
\tilde W(f^\ast g) - \tilde W(g) = \tilde S(f; g)
\end{equation}
 It's ease to see that from \r{23}, \r{44} and from
\begin{equation}\label{46}
\ln{\sqrt{f^\ast g(x)}} = \ln{\sqrt{g(f)}} + \ln{\Delta^{f}_{x}}
\end{equation}
where
\begin{equation}\label{47}
\Delta^{f}_{x} = \det\frac{\partial f^{\alpha}(x)}{\partial x^{\beta}}
\end{equation}
 we obtain
\begin{equation}\label{48}
\tilde S(f; g) = S(\sigma ; g_{\alpha\beta} / (\sqrt{g})^{1/k})
\bigg\vert_{\sigma = \frac{1}{k} \ln\Delta^{f^{-1}}_{x}}
\end{equation}
This action has a cocyclic property for the $Diff(2k)$ group
\begin{equation}\label{49}
\tilde S(f\circ h; g) =\tilde S(f; h^\ast g) + \tilde S(h; g)
\end{equation}
Therefore we can define the change of the measure for $Diff(2k)$
transformation in the case when we are using Weyl-invariant regularization:
\begin{equation}\label{50}
 D_{f^\ast g} \varphi = D_{g} \varphi \exp{\tilde S(f; g_{\alpha\beta})}
\end{equation}
where  $\tilde S(f; g)$ is defined in \r{48} and has the property of being
1-cocycle of $Diff(2k)$ group.
Finally we can substitute \r{32} and \r{41} in \r{48} and obtain the
$Diff(4), Diff(6)$ cocycles
\begin{equation}\label{51}
\tilde {\bf S}_{2}(f; g) = {\bf S}_{2}(\sigma ; g_{\alpha\beta}
/ (\sqrt{g})^{1/2})
\bigg\vert_{\sigma = \frac{1}{k} \ln\Delta^{f^{-1}}_{x}}\quad k=2,3
\end{equation}
The corresponding $Diff(2)$ cocycle has been obtained in Ref. \cite{KMM}.

\section{\bf Conclusions and outlook}

In this work we discussed the consequencies of the Wess-Zumino consistency
condition \cite{WZ} on the structure of trace anomaly in any even dimensions.
The Wess-Zumino condition follows from the fact that the second Weyl
variation of the
effective action $W(g)$ has to be symmetric, that means, that the first
Weyl variation of the anomaly also has to be symmetric. The variation
of the Euler density  satisfies this condition in any even dimensions
due to Bianchi identity. Beside that
Weyl invariant terms (like $C^{2}$ in $d=4$)\cite{DS} also, evidently,
satisfy the WZ consisitency condition. The other terms,
with lower order on R can be obtained as variations of local functionals
and can be removed by adding local counterterms to an effective action.
This last statement, which is proved exactly at d=2,4,6, and remains
a hypothesis at higher $d$, completes the classification of the structure
of trace anomaly in all dimensions, following from the WZ consistency
condition, and that classification coincides with that of Deser and Shwimmer
\cite{DS}.

Next, we have considered the responce of the functional integral on the
finite Weyl transformation, which, as is well-known for all the anomalous
symmetries, is connected to the cocycles of the corresponding group.
The corresponding cocycle of the Weyl group is calculated explicitly
for d=6 and it is shown, that, surprisingly, it can be brought to the
form, which is only of a second order over the  Weyl parameter $\sigma$.
This form contains a Weyl-invariant differential operators of the sixth order.
These results give rise to the general hypothesis, concerning the
cocycles of the Weyl group in an arbitrary dimensions - it appear
to be the exact connection between Euler class, cocycle of the Weyl group,
and the Weyl-invariant differential operator of order $d$
(dimensionality of the space-time) and weight zero.

Finally, it is shown, how one can use the cocycle of the Weyl group
for obtaining the cocycle of the diffeomorphisms group in all dimensions,
and the exact expressions for this transformation are obtained.

{\bf{Acknowledgments}}

This work was supported in part by the INTAS grant \# 93-2058, DRK
and RPM are also supported by the ISF grant \# RYU000.

\end{document}